\documentclass[12pt]{iopart}
\usepackage{graphicx}

\usepackage{iopams}

\begin{document}
\title[]{The Work Done by an External Field}
\author{Mingliang Zhang and D. A. Drabold}
\address{Department of Physics and Astronomy, Ohio University, Athens, Ohio 45701}
\eads{\mailto{zhangm@ohio.edu},~\mailto{drabold@ohio.edu}}
\begin{abstract}

From the change in kinetic energy induced by an external field, we discuss
the applicable conditions for the Mott-Davis
and Moseley-Lukes form of the Kubo-Greenwood formula (KGF) for the electrical conductivity which
has been implemented in \textit{ab initio} codes. We show that the simplified
KGF is suitable only for computing the AC conductivity at sufficiently high frequency and
when the gradient of the carrier density is small.
\end{abstract}

\maketitle

\section{Introduction}

\label{ind}

The Kubo-Greenwood formula (KGF) is widely adopted to compute AC conductivity
for amorphous semiconductors and to estimate the DC
conductivity from an extrapolation procedure\cite{motda,gal}. Greenwood\cite{Gre} based his derivation on a
kinetic expression for current density which is only justified when the
gradient of the carrier concentration is small\cite{sg}. Mott and
Davis\cite{motda}, Moseley and Lukes\cite{mos} developed a simplified
version of KGF without using an explicit expression for the current density, but
they made two implicit assumptions: (1) the Joule heat produced by a sample
per unit time is $\Omega\mathbf{j}\cdot\mathbf{E}$, where $\Omega$ is the
volume of sample, $\mathbf{E}$ the strength of external electric field,
$\mathbf{j}$ is the current density; and (2) identifying the power
$\Omega\mathbf{j}\cdot\mathbf{E}$ of current as the absorption energy per unit
time $\Gamma$ from AC field:
\begin{equation}
\Gamma=\sum_{fi}\hbar\omega_{fi}(w_{fi}P_{i}-w_{if}P_{f}), \label{abs}%
\end{equation}
where $P_{i}$ is the occupation probability of the initial state $|i\rangle$,
$w_{fi}$ is the transition probability per unit time from initial state
$|i\rangle$ to final state $|f\rangle$, $\hbar\omega_{fi}$ is the energy
difference between $|f\rangle$ and $|i\rangle$. The conductivity is read off
from
\begin{equation}
\frac{\Omega}{2}\sum_{\alpha,\beta}\sigma_{\alpha\beta}E_{\alpha}E_{\beta}%
=\Gamma,~~\alpha,\beta=x,y,z, \label{conm}%
\end{equation}
where the factor $1/2$ in the L.H.S comes from averaging the power of current
over one period of the AC field. The AC conductivity at frequency $\omega$
is\cite{motda,mos}%
\begin{equation}
\sigma(\omega)=\frac{2\pi e^{2}\hbar^{3}\Omega}{m^{2}}\int dE\frac
{[f(E)-f(E+\hbar\omega)]|D|_{av}^{2}N(E)N(E+\hbar\omega)}{\hbar\omega},
\label{acc}%
\end{equation}
where $\Omega$ is the volume of sample, $f$ is the Fermi distribution
function, $N(E)$ is the density of states, $D_{E^{\prime},E}=\int d^{3}%
x\psi_{E^{\prime}}^{\ast}\partial\psi_{E}/\partial x$, and `av' represents an
average over all states having energy near $E^{\prime}=E+\hbar\omega$. The DC
conductivity is obtained by taking the $\omega\rightarrow0$ limit in
(\ref{acc})\cite{motda}:
\begin{equation}
\sigma_{dc}=\frac{2\pi e^{2}\hbar^{3}\Omega}{m^{2}}\int dE|D|_{av}%
^{2}[N(E)]^{2}\frac{df}{dE}. \label{dcc}%
\end{equation}
Since $w_{fi}$ contains an energy-conserving delta function $\delta(E_{f}%
-E_{i}-\hbar\omega)$, the assumption (2) means that the system is driven by a
radiation field with frequency $\omega=\omega_{fi}$. The photon energy for a
DC field is zero, so that the absorbed energy of the system from a DC field
should vanish. According to (\ref{conm}), $\sigma_{dc}$ would be zero.
However a DC voltage does produce Joule heat in a conductor or a semiconductor.
One may conclude that (i) (\ref{acc}) is not suited to AC fields with
low frequencies; and (ii) (\ref{dcc}) is not consistent with (\ref{conm}),
although the limiting procedure from (\ref{acc}) to (\ref{dcc}) seems legitimate.
On the other hand, (\ref{acc}) has been used to compute AC\ conductivity and
extrapolate $\sigma_{dc}$ in liquid carbon\cite{gal}. (\ref{dcc}) is implemented in
SIESTA\cite{abt} to calculate $\sigma_{dc}$ for a-Si and a-Si:H. The DC
conductivities obtained fall in the range of observed values in different
systems. Therefore extrapolation from (\ref{acc}) and (\ref{dcc}) must
represent the correct $\sigma_{dc}$ to some extent.

The aim of this paper is to explore and resolve these controversial issues.
In Sec.\ref{wep}, we derive the rate $dK/dt$ of change kinetic energy from the
time-dependent Schr\"{o}dinger equation. This rigorous expression of power is
applicable to an arbitrary electromagnetic field, and leads to a proper current
density\cite{short,4c,sg}. Using the approximation implied by assumption (1),
the rigorous current density reduces to the kinetic expression used by
Greenwood\cite{Gre}. Thus the simplified derivations\cite{motda,mos} suffer
the same error as the original one: the formula is suitable only when the gradient of the
carrier density is small, a condition derived from the current
operator\cite{sg}.

To find the connection between $dK/dt$ (well defined for any electromagnetic
field) and $\Gamma$ which is only defined for a radiation field, we calculate
$dK/dt$ for three successively more complex cases: DC voltage, AC voltage and
arbitrary field. For both DC and low frequency AC voltages (i.e. any field
described by a scalar potential), $dK/dt$ comes from the work done by the
internal force. When a time-dependent vector potential appears, there is an additional $\partial
K/\partial t$ term contributing to the change in kinetic energy which results
from the time dependence of vector potential $\partial {\bf{A}}/\partial t$.

A DC voltage described by a time-independent scalar potential can be described
from both stationary perturbation theory (SPT) and time-dependent theory
(TDPT). Sec.\ref{si} will show that in TDPT only when we introduce the
interaction with the time-independent scalar potential in a specified way
(physically reasonable), are the descriptions of states and $dK/dt$ in TDPT
consistent with those in SPT. This is a natural requirement for a
self-consistent theory. In Sec.\ref{psi}, we show that the Joule heat
from a low frequency AC field described by a time-dependent scalar potential
is the work done by the internal force. As expected, when $\omega=0$, the
result for AC voltage reduces to the DC result.
This is possible because we consistently use a scalar potential to describe the
interaction with a low frequency external field. If we adopt a different
gauge, both scalar potential and vector potential are needed. In Sec.\ref{em},
we calculate the rate of change of the kinetic energy induced by a general
electromagnetic field described by both vector and scalar potentials.
We will see that assumption (2) renders the simplified KGF\cite{motda,mos} (\ref{acc}) only suitable for computing conductivity at sufficiently high frequency,
i.e for a radiation field. This is the origin of the confusion: viewing
$\Gamma$ as the unique reason for change in kinetic energy\cite{motda,mos}
$dK/dt$ and taking the $\omega\rightarrow0$ limit in (\ref{acc})\cite{motda} are not valid. Our formulation is applicable to an arbitrary electromagnetic field in
any gauge: applying Greenwood's gauge $\mathbf{A}=-\mathbf{E}t$\ for a DC
voltage in the $\partial K/\partial t$ term of our strict formula does lead to
(\ref{dcc}). However (\ref{dcc}) neglected the contribution from the work done
by the internal force.

\section{Rate of change in kinetic energy}

\label{wep}

We will use the Schr\"{o}dinger picture. Consider a system with $N$ electrons
+ $\mathcal{N}$ nuclei in an external electromagnetic field $(\mathbf{A}%
,\phi)$, at time $t$, and the state of system is described by $\Psi^{\prime
}(\mathbf{r}_{1},\mathbf{r}_{2},\cdots,\mathbf{r}_{N};t)$. To save space, we
will not write out the nuclear coordinates explicitly. $\Psi^{\prime}$
satisfies the Schr\"{o}dinger equation%
\begin{equation}
i\hbar\partial\Psi^{\prime}/\partial t=H^{\prime}\Psi^{\prime}, \label{tds1}%
\end{equation}
where $H^{\prime}=H+H_{mf}(t)$, $H_{mf}(t)$ is the interaction between system
and external field $(\mathbf{A},\phi)$. The time dependence of $H_{mf}(t)$
comes from the external field. $H$ is the Hamiltonian of the system without
external field: $H|m\rangle=\varepsilon_{m}|m\rangle$, we use $|m\rangle$ or $\Psi_{m}$
to denote the $m^{th}$ stationary state of the system. If the system is in a
thermal bath at temperature $T$, before introducing external field, the
probability that the system is in state $\Psi_{m}$ is $P_{m}=e^{-\beta \varepsilon_{m}%
}/Z$, where $Z=\sum_{n}e^{-\beta \varepsilon_{n}}$ is the partition function.

For a system in external field $(\mathbf{A},\phi)$, the velocity operator
$\mathbf{v}_{i}$ for the $i^{th}$ electron is\cite{lv3} $\mathbf{v}_{i}%
=m^{-1}\mathbf{p}_{i}^{m}$, where $\mathbf{p}_{i}^{m}=-i\hbar\nabla
_{\mathbf{r}_{i}}-e\mathbf{A}(\mathbf{r}_{i};t)$ and $\mathbf{r}_{i}$ are the
mechanical momentum and position operators of the $i^{th}$ electron, $e=-1.6\times 10^{-19}$C is
the charge of electron. Similarly $\mathbf{v}_{\alpha}=$ $\mathbf{p}_{\alpha
}^{m}/M_{\alpha}$ is the velocity of the $\alpha^{th}$ nucleus, $\mathbf{p}%
_{\alpha}^{m}=-i\hbar\nabla_{\mathbf{R}_{\alpha}}+Z_{\alpha}e\mathbf{A}%
(\mathbf{R}_{\alpha};t)$ and $\mathbf{R}_{\alpha}$ are the mechanical momentum
and position operators of the $\alpha^{th}$ nucleus, $-Z_{\alpha}e$ is the
charge of the $\alpha^{th}$ nucleus. The average kinetic energy of the system in
state $\Psi^{\prime}(t)$ is%
\begin{equation}
K_{\Psi^{\prime}}(t)=\int d\tau\Psi^{\prime\ast}(t)\widehat{K}(t)\Psi^{\prime
}(t), \label{avk}%
\end{equation}
where $d\tau=d\mathbf{r}_{1}d\tau^{\prime}$, $d\tau^{\prime}=d\mathbf{r}%
_{2}\cdots d\mathbf{r}_{N}$, the arguments of $\Psi^{\prime}$ are
$(\mathbf{r}_{1},\mathbf{r}_{2},\cdots,\mathbf{r}_{N};t)$, and
\begin{equation}
\widehat{K}(t)=\sum_{i}\frac{(\mathbf{p}_{i}^{m})^{2}}{2m}+\sum_{\alpha}%
\frac{(\mathbf{p}_{\alpha}^{m})^{2}}{2M_{\alpha}} \label{ko}%
\end{equation}
is the kinetic energy operator of the whole system in an external field. The
time dependence in $\widehat{K}(t)$ arises from that of $\mathbf{A}$. To
compute the macroscopic response to a mechanical perturbation, the
coarse-grained average and ensemble average can be done in the final
stage\cite{short,4c,sg}. In this paper we only discuss the average over the
state of the system.

With the help of (\ref{tds1}), the rate of change in the average kinetic energy
is\cite{lv3}%
\begin{equation}
\frac{d}{dt}K_{\Psi^{\prime}}(t)=\frac{1}{i\hbar}\int d\tau\Psi^{\prime\ast
}(t)[\widehat{K},H^{\prime}]\Psi^{\prime}(t)
+\int d\tau\Psi^{\prime\ast}(t)\frac{\partial\widehat{K}(t)}{\partial t}%
\Psi^{\prime}(t).\label{rat}
\end{equation}%
The Hamiltonian $H^{\prime}$ of system in external field can be written as%
\begin{equation}
H^{\prime}=\widehat{K}(t)+V_{\phi}+H_{2}, \label{hp1}%
\end{equation}
where%
\begin{equation}
V_{\phi}=\sum_{i}e\phi(\mathbf{r}_{i};t)-\sum_{\alpha}Z_{\alpha}%
e\phi(\mathbf{R}_{\alpha};t), \label{vp}%
\end{equation}
is the potential energy of the system in an external field $(\mathbf{A},\phi)$, and
\begin{equation}
H_{2}=\frac{1}{2}\sum_{ij}V_{2}(\mathbf{r}_{i},\mathbf{r}_{j})+\sum_{i\alpha
}V_{1}(\mathbf{r}_{i},\mathbf{R}_{\alpha})
+\frac{1}{2}\sum_{\alpha\beta}U_{2}(\mathbf{R}_{\alpha},\mathbf{R}_{\beta}),
\label{h2}%
\end{equation}%
represents the internal interactions of the system, $V_{1}(\mathbf{r}%
_{i},\mathbf{R}_{\alpha})$ is the interaction energy of the $i^{th}$ electron
and $\alpha^{th}$ nucleus, $V_{2}(\mathbf{r}_{i},\mathbf{r}_{j})$ is the
interaction energy between the $i^{th}$ electron and the $j^{th}$ electron,
$U_{2}(\mathbf{R}_{\alpha},\mathbf{R}_{\beta})$ is the interaction energy
between the $\alpha^{th}$ and the $\beta^{th}$ nuclei. The 1$^{st}$ term in
(\ref{rat}) can be changed to:%
\begin{equation}
\lbrack\widehat{K},H^{\prime}]=[\widehat{K},H_{2}]+[\widehat{K},V_{\phi}].
\label{dy}%
\end{equation}
Here $(i\hbar)^{-1}[\widehat{K},H_{2}]$ represents the power due to the internal
force:%
\begin{equation}
\frac{1}{i\hbar}[\widehat{K},H_{2}]=\sum_{i}\{\mathbf{f}_{i}\cdot
\mathbf{v}_{i}-\frac{i\hbar}{2m}\nabla_{\mathbf{r}_{i}}\cdot\mathbf{f}_{i}\}
+\sum_{\alpha}\{\mathbf{f}_{\alpha}\cdot\mathbf{v}_{\alpha}-\frac{i\hbar
}{2M_{\alpha}}\nabla_{\mathbf{R}_{\alpha}}\cdot\mathbf{f}_{\alpha}\},
\label{pinf}%
\end{equation}
where%
\begin{equation}
\mathbf{f}_{\alpha}=-\{\nabla_{\mathbf{R}_{\alpha}}[\sum_{i}V_{1}%
(\mathbf{r}_{i},\mathbf{R}_{\alpha})+\frac{1}{2}\sum_{\beta}U_{2}%
(\mathbf{R}_{\alpha},\mathbf{R}_{\beta})]\}, \label{fh}%
\end{equation}
is the internal force on the $\alpha^{th}$ nucleus, and%
\begin{equation}
\mathbf{f}_{i}=-\{\nabla_{\mathbf{r}_{i}}[\frac{1}{2}\sum_{j}V_{2}%
(\mathbf{r}_{i},\mathbf{r}_{j})+\sum_{\alpha}V_{1}(\mathbf{r}_{i}%
,\mathbf{R}_{\alpha})]\}, \label{fe1}%
\end{equation}
is the internal force on the $i^{th}$ electron. From now on we will not write
out the corresponding terms for nuclei which are similar to those for
electrons. The second term of (\ref{dy}) represents the power due to the electric field described by the scalar potential:
\begin{equation}
\frac{1}{i\hbar}[\widehat{K},V_{\phi}]=\sum_{i}e[-\nabla_{\mathbf{r}_{i}}%
\phi(\mathbf{r}_{i};t)]\cdot\mathbf{v}_{i}+\sum_{i}\frac{e}{2m}i\hbar
\lbrack\nabla_{\mathbf{r}_{i}}\cdot\nabla_{\mathbf{r}_{i}}\phi(\mathbf{r}%
_{i};t)]. \label{cKPE}%
\end{equation}

To calculate the 2$^{nd}$ term in (\ref{rat}), one should notice that
$[\mathbf{p}_{i}^{m},\partial\mathbf{p}_{i}^{m}/\partial t]\neq0$ and
$[\mathbf{p}_{\alpha}^{m},\partial\mathbf{p}_{\alpha}^{m}/\partial t]\neq0$:%

\begin{equation}
\frac{\partial\widehat{K}(t)}{\partial t}=\sum_{i}[-e\frac{\partial
\mathbf{A}(\mathbf{r}_{i};t)}{\partial t}]\cdot\mathbf{v}_{i}+\sum_{i}\frac
{1}{2m}i\hbar e[\nabla_{\mathbf{r}_{i}}\frac{\partial\mathbf{A}(\mathbf{r}%
_{i};t)}{\partial t}]. \label{pkpt}%
\end{equation}
Substituting (\ref{pinf},\ref{cKPE},\ref{pkpt}) into (\ref{rat}), one finds%
\begin{equation}
\frac{d}{dt}K_{\Psi^{\prime}}(t)=\int d\tau\Psi^{\prime\ast}(t)\sum
_{i}e\mathbf{E}(\mathbf{r}_{i};t)\cdot\mathbf{v}_{i}\Psi^{\prime}(t)
\label{rk1}%
\end{equation}%
\[
+\int d\tau\sum_{i}\frac{1}{2m}[e\mathbf{E}(\mathbf{r}_{i};t)]\cdot\lbrack
i\hbar\nabla_{\mathbf{r}_{i}}\Psi^{\prime}(t)\Psi^{\prime\ast}(t)]
\]%
\[
+\int d\tau\Psi^{\prime\ast}(t)\sum_{i}\mathbf{f}_{i}\cdot\mathbf{v}_{i}%
\Psi^{\prime}(t)+\int d\tau\frac{1}{2m}\sum_{i}\mathbf{f}_{i}\cdot\lbrack
i\hbar\nabla_{r_{i}}\Psi^{\prime}(t)\Psi^{\prime\ast}(t)],
\]
where $\mathbf{E}(\mathbf{r}_{i};t)=-\nabla\phi(\mathbf{r}_{i};t)-\partial
\mathbf{A}(\mathbf{r}_{i};t)/\partial t$ is the electric field at
$\mathbf{r}_{i}$. To obtain the 2$^{nd}$ and 4$^{th}$ terms in (\ref{rk1}), we integrated by parts. (\ref{rk1}) is a form of the Ehrenfest theorem: the rate
of change in kinetic energy equals the work done per unit time by the external
electric field (the first two terms) and the internal force (the last two
terms). The 1$^{st}$ and 3$^{rd}$ terms are the corresponding quantum average
values of the powers in classical mechanics. The 2$^{nd}$ and 4$^{th}$ terms
are produced by the commutation relation between momentum and position: they
will disappear in classical limit.

In the first two terms of (\ref{rk1}), exchanging the integration 
variables $\mathbf{r}_{k}(k=2,3,\cdots N)\leftrightarrow\mathbf{r}_{1}$, using
the antisymmetry of the many-electron wave function and changing 
 $\mathbf{r}_{1}$ to $\mathbf{r}$, they become $\int_{\Omega
}d\mathbf{r}$ $[\mathbf{E}(\mathbf{r};t)]\cdot\mathbf{j}_{m}(\mathbf{r};t)$:%
\begin{equation}
\mathbf{j}_{m}(\mathbf{r};t)=Ne\int d\tau^{\prime}\Psi^{\prime\ast}%
\mathbf{v}\Psi^{\prime}+\frac{i\hbar e}{2m}\nabla_{\mathbf{r}}n^{\prime
}(\mathbf{r};t), \label{trc1}%
\end{equation}
where the arguments of $\Psi^{\prime}$ are $(\mathbf{r,r}_{2}\mathbf{r}%
_{3}\cdots\mathbf{r}_{N};t)$, $n^{\prime}(\mathbf{r};t)=N\int d\tau^{\prime
}\Psi^{\prime}\Psi^{\prime\ast}$ is the carrier density. $\mathbf{j}_{m}$
defined by (\ref{trc1}) is the same as the rigorous microscopic current
density obtained from the microscopic response method\cite{short} and
polarization density\cite{4c}. Only when the gradient of the carrier density is
small, can one neglect the second term in (\ref{trc1}) and replace
$\int_{\Omega}d\mathbf{r}$ $\mathbf{E}\cdot\mathbf{j}_{m}$with $\Omega
\mathbf{j}\cdot\mathbf{E}$. By means of the equivalence between the
microscopic response method and the Kubo formula\cite{sg}, the first term in
(\ref{trc1}) corresponds to the kinetic expression $e$Tr[$\rho^{\prime
}(t)\mathbf{v}]$ of Greenwood\cite{Gre}, where $\rho^{\prime}(t)$ is the
density matrix of the system in external field. Thus assumption (1) is equivalent
to using $e$Tr[$\rho^{\prime}(t)\mathbf{v}]$, the kinetic expression for the current density.

\section{DC voltage}

\label{si}

Let us adopt the common gauge in which a DC field is solely described by a
time-independent scalar potential. In this gauge, both stationary
perturbation theory (SPT) and time-dependent perturbation theory (TDPT)
can be used, and they should give the same results for any observable
quantities. As we will see, the first nonzero contribution to the rate of
change of kinetic energy is second order in $V$ (consistent with our
macroscopic experience), so to formulate a consistent approximation, we
will carry out perturbation theory to second order in $V$.

Applying a DC voltage on a piece of conductor or semiconductor, after a short
transient period, the system will evolve to a steady state if the system is in good thermal contact with the environment such that the Joule heat evolved can be
completely removed from the system. The constant external voltage establishes
a time-independent electric field \textit{inside} the system. The system is
described by a Hamiltonian $H^{\prime}=H+V$, where $H$ is the Hamiltonian of
system without external DC voltage, $V$ is given by (\ref{vp}) with a steady
scalar potential $\phi$.

If the system is initially in an eigenstate $\Psi_{j}$ of $H$ with eigenvalue
$\varepsilon_{j}$, after the short transient period, the system will be in the
stationary state $\Psi_{j}^{\prime}$ of $H^{\prime}$ with eigenvalue
$\varepsilon_{j}^{\prime}$. Denote $V_{kj}=\langle\Psi_{k}|V|\Psi_{j}\rangle$,
then one can easily compute\cite{lv3} $\Psi_{j}^{\prime}$ and $\varepsilon
_{j}^{\prime}$ to second order in $V$:%
\begin{equation}
\Psi_{j}^{\prime}=\Psi_{j}+\sum_{p=1,2}\{\sum_{m(\neq j)}c_{mj}^{(p)}\Psi
_{m}+c_{jj}^{(p)}\Psi_{j}\}. \label{spt}%
\end{equation}
Since $H^{\prime}$ is time-independent, the time evolution of system is given by%
\begin{equation}
\Psi_{j}^{\prime}(t)=e^{-it\varepsilon_{j}^{\prime}/\hbar}\Psi_{j}^{\prime}.
\label{sta}%
\end{equation}
Substituting $\Psi_{j}^{\prime}$ and $\varepsilon_{j}^{\prime}$ obtained from
SPT into (\ref{sta}), to second order in $V$,%
\begin{equation}
\Psi_{j}^{\prime}(t)=e^{-it\varepsilon_{j}/\hbar}\Psi_{j}%
+e^{-it\varepsilon_{j}/\hbar}\sum_{p=1,2}\{\sum_{m(\neq j)}b_{mj}^{(p)}%
\Psi_{m}+b_{jj}^{(p)}\Psi_{j}\}, \label{sta1}%
\end{equation}
where%
\begin{equation}
b_{mj}^{(1)}=\frac{V_{mj}}{\varepsilon_{j}-\varepsilon_{m}}~~ for m\neq
j,~~b_{jj}^{(1)}=-\frac{it}{\hbar}V_{jj} \label{ce1}%
\end{equation}
and%
\begin{equation}
b_{mj}^{(2)}=\sum_{k(\neq j)}\frac{V_{mk}V_{kj}}{\left(  \varepsilon
_{j}-\varepsilon_{k}\right)  \left(  \varepsilon_{j}-\varepsilon_{m}\right)  }
\label{ce2}%
\end{equation}%
\[
-V_{jj}\frac{V_{mj}}{\left(  \varepsilon_{j}-\varepsilon_{m}\right)  ^{2}%
}-\frac{it}{\hbar}V_{jj}\frac{V_{mj}}{\varepsilon_{j}-\varepsilon_{m}}~~
for~~ m\neq j,%
\]
and%
\begin{equation}
b_{jj}^{(2)}=-\frac{1}{2}\sum_{k(\neq j)}\frac{|V_{kj}|^{2}}{\left(
\varepsilon_{j}-\varepsilon_{k}\right)  ^{2}}-\frac{it}{\hbar}\sum_{k(\neq
j)}\frac{|V_{kj}|^{2}}{\varepsilon_{j}-\varepsilon_{k}}-\frac{t^{2}V_{jj}^{2}%
}{2\hbar^{2}}. \label{ce2d}%
\end{equation}

Now consider the viewpoint of TDPT, in which the perturbation $V$ causes
transitions from $\Psi_{j}$ to other eigenstates $\Psi_{k}$ of $H$. Using the
familiar expansion\cite{lv3}
\begin{equation}
\Psi_{j}^{\prime}(t)=\Psi_{j}e^{-it\varepsilon_{j}/\hbar}
+\sum_{p=1,2}[\sum_{m(\neq j)}a_{mj}^{(p)}(t)e^{-it\varepsilon_{m}/\hbar}%
\Psi_{m}+a_{jj}^{(p)}(t)e^{-it\varepsilon_{j}/\hbar}\Psi_{j}],\label{exp}%
\end{equation}%
the first order expansion coefficients satisfy%
\begin{equation}
i\hbar\frac{da_{mj}^{(1)}(t)}{dt}=e^{it(\varepsilon_{m}-\varepsilon_{j}%
)/\hbar}V_{mj}~~ for~~ m\neq j, \label{b1}%
\end{equation}
and%
\begin{equation}
i\hbar\frac{da_{jj}^{(1)}(t)}{dt}=V_{jj}. \label{b1d}%
\end{equation}
To make (\ref{exp}) consistent with (\ref{sta1}) at order $V$,\ we have to
integrate (\ref{b1}) by adiabatically introducing the interaction
$\int_{-\infty}^{t}dt^{\prime}e^{\lambda t^{\prime}}$ ($\lambda\rightarrow
0^{+}$) and integrate (\ref{b1d}) by $\int_{0}^{t}dt^{\prime}$.
This procedure is reasonable because the system is initially in state
$\Psi_{j}$, the transition from $\Psi_{j}$ to another state $\Psi_{m}(m\neq
j)$ requires some time. On the other hand, the probability amplitude
 of state $\Psi_{j}$ begins to decrease immediately.

The second order expansion coefficients satisfy%
\begin{equation}
i\hbar\frac{da_{mj}^{(2)}(t)}{dt}=\sum_{k(\neq j)}a_{kj}^{(1)}%
(t)e^{it(\varepsilon_{m}-\varepsilon_{k})/\hbar}V_{mk}
+a_{jj}^{(1)}(t)e^{it(\varepsilon_{m}-\varepsilon_{j})/\hbar}V_{mj},\label{a2}%
\end{equation}%
and%
\begin{equation}
i\hbar\frac{da_{jj}^{(2)}(t)}{dt}=\sum_{k(\neq j)}a_{kj}^{(1)}%
(t)e^{it(\varepsilon_{j}-\varepsilon_{k})/\hbar}V_{jk}
+a_{jj}^{(1)}(t)V_{jj}.\label{a2d}%
\end{equation}%
If we integrate (\ref{a2}) by adiabatically introducing interaction
$\int_{-\infty}^{t}dt^{\prime}e^{\lambda t^{\prime}}$ ($\lambda\rightarrow
0^{+}$), integrate (\ref{a2d}) by $\int_{0}^{t}dt^{\prime}$ and in
$a_{jj}^{(2)}$ drop one term with wrong time factor $e^{-it\varepsilon
_{k}/\hbar}$, we almost reproduce (\ref{sta1}) except without the $1/2$ factor
in the first term of $b_{jj}^{(2)}$.

There are two differences between SPT and TDPT of concern to us here. In SPT no
equation exists to determine $c_{jj}^{(p)}(p=1,2)$ in (\ref{spt}). The
perturbed wave function (\ref{spt}) is not normalized, if one does not include
$\sum_{p=1,2}c_{jj}^{(p)}\Psi_{j}$. $c_{jj}^{(p)}(p=1,2)$ are determined from
the normalization of $\Psi_{j}^{\prime}$ to the corresponding order\cite{lv3}.
In TDPT, $a_{jj}^{(p)}(t)$ $(p=1,2)$ are determined by (\ref{b1d},\ref{a2d}).
Using (\ref{tds1}), one can easily find $(d/dt)[\int d\tau\Psi_{j}^{\prime
\ast}(t)\Psi_{j}^{\prime}(t)]=0$. Thus the perturbed wave function (\ref{exp})
is normalized if $\int d\tau\Psi_{j}^{\ast}\Psi_{j}=1$. If in (\ref{exp}), we
only considered $a_{mj}^{(1)}(t)$ for $m\neq j$, and used the normalization
condition of $\Psi_{j}^{\prime}(t)$ to determine $a_{jj}^{(p)}(t)$ $(p=1,2)$,
we would not reproduce (\ref{sta1}). Therefore the suggested means of introducing the
interaction is necessary to make TDPT consistent with SPT.

When the DC field is described by a time-independent scalar potential, the
rate of change in kinetic energy can be written as%
\begin{equation}
\frac{d}{dt}K_{\Psi_{j}^{\prime}(t)}=\frac{1}{i\hbar}\int d\tau\Psi
_{j}^{\prime\ast}(t)[K_{0},H]\Psi_{j}^{\prime}(t)
+\frac{1}{i\hbar}\int d\tau\Psi_{j}^{\prime\ast}(t)[H,V]\Psi_{j}^{\prime}(t),\label{kcs}%
\end{equation}%
where $K_{0}$ is the kinetic energy when vector potential is zero. Because
$[K_{0},H]=[K_{0},H_{2}]$ and $[H,V]=[K_{0},V]$, after a comparison with
(\ref{pinf},\ref{cKPE}), one may say that the first term in (\ref{kcs}) is the
power due to the internal force, and the second term in (\ref{kcs}) is the
power due to the external force. Of course, the effect of external field is
also reflected in $\Psi_{j}^{\prime}(t)$. With the help of (\ref{exp}), one
can easily show that to order V$^{2}$, the second term in (\ref{kcs}) is zero.
(\ref{kcs}) becomes%
\begin{equation}
\frac{d}{dt}K_{\Psi_{j}^{\prime}(t)}=\frac{1}{i\hbar}\int d\tau\Psi
_{j}^{\prime\ast}(t)[K_{0},H]\Psi_{j}^{\prime}(t) \label{inw1}%
\end{equation}%
\[
=\frac{1}{i\hbar}\sum_{kl}a_{kj}^{(1)\ast}a_{lj}^{(1)}[\varepsilon_{l}-\varepsilon_{k}%
]K_{0kl}
+\frac{1}{i\hbar}\sum_{l}[\varepsilon_{l}-\varepsilon_{j}][a_{lj}^{(2)}K_{0jl}%
-a_{lj}^{(2)\ast}K_{0lj}],
\]
where $K_{0jl}=\langle j|K_{0}|l\rangle$ and $a_{lj}^{(p)}=a_{lj}%
^{(p)}(t)e^{i(\varepsilon_{j}-\varepsilon_{l})t/\hbar}$ $(p=1,2)$. The change
in kinetic energy is produced by the power of the internal force. It is easy
to check that without an external field, the internal force does no work, the
average kinetic energy $K_{\Psi_{j}}(t)=\int d\tau\Psi_{j}^{\ast}(t)K_{0}%
\Psi_{j}(t)$ in stationary state $\Psi_{j}(t)$ does not change with time:%
\begin{equation}
\frac{d}{dt}K_{\Psi_{j}}(t)=\frac{1}{i\hbar}\int d\tau\Psi_{j}^{\ast}%
(t)[K_{0},H]\Psi_{j}(t)=0. \label{we}%
\end{equation}
The Joule heat comes from a steady voltage which changes the state of system.

\section{Low frequency AC voltage}

\label{psi}

We will use a gauge in which a low frequency AC field is solely described by a
time-dependent scalar potential. The interaction of system with an AC voltage
is given by (\ref{vp}) with a periodic scalar potential $\phi$:%

\begin{equation}
V(t)=Fe^{-i\omega t}+F^{\dagger}e^{i\omega t}. \label{ac}%
\end{equation}
An AC voltage will produce a time-dependent current. According to Ampere's law,
the time-dependent current will produce a time-dependent magnetic induction.
Therefore a time-dependent vector potential $\mathbf{A}(\mathbf{r},t)$ must
accompany the AC voltage. (\ref{ac}) is only suitable for low frequency
$\omega\ll\sigma/\epsilon_{0}$, where $\sigma$ is the DC conductivity for the system.

When the frequency of an AC voltage approaches zero, its properties should be
the same as those of DC voltage. Therefore we must integrate the equations of
probability amplitudes $a_{lj}^{(p)}(t)$ $(p=1,2)$ for $V(t)$ in the same way
as those for the DC voltage:%
\begin{equation}
a_{lj}^{(1)}(t)=-\frac{F_{lj}e^{i(\omega_{lj}-\omega)t}}{\hbar(\omega
_{lj}-\omega)}-\frac{F_{jl}^{\ast}e^{i(\omega_{lj}+\omega)t}}{\hbar
(\omega_{lj}+\omega)}~~ for~~ l\neq j, \label{1vt}%
\end{equation}
and%
\begin{equation}
a_{jj}^{(1)}(t)=\frac{F_{jj}(e^{-i\omega t}-1)}{\hbar\omega}-\frac
{F_{jj}^{\ast}(e^{i\omega t}-1)}{\hbar\omega}. \label{1vtd}%
\end{equation}
We will not write down the expressions for $a_{lj}^{(2)}(t)$ and $a_{jj}%
^{(2)}(t)$, they are too long. To the 2$^{nd}$ order in $V(t)$, the state of system is%
\begin{equation}
\Psi_{j}^{\prime}(t)=\Psi_{j}e^{-it\varepsilon_{j}/\hbar}+\sum_{p=1,2}\sum
_{l}a_{lj}^{(p)}(t)\Psi_{l}e^{-it\varepsilon_{l}/\hbar}. \label{svt}%
\end{equation}
The rate of change in kinetic energy is%
\begin{equation}
\frac{d}{dt}K_{\Psi_{j}^{\prime}}(t)=\frac{1}{i\hbar}\int d\tau\Psi
_{j}^{\prime\ast}(t)[K_{0},H]\Psi_{j}^{\prime}(t)
+\frac{1}{i\hbar}\int d\tau\Psi_{j}^{\prime\ast}(t)[H,V(t)]\Psi_{j}^{\prime
}(t), \label{en}%
\end{equation}
For an AC voltage, to obtain the dissipated energy, we must average (\ref{en})
over a period $T=2\pi/\omega$ of the AC voltage\cite{lv8}: $T^{-1}\int_{0}%
^{T}dt$. Using $a_{lj}^{(p)}(t)$ and $a_{jj}^{(p)}(t)$ $(p=1,2)$, one can show
that to 2$^{nd}$ order in $V(t)$,%
\begin{equation}
T^{-1}\int_{0}^{T}dt\frac{1}{i\hbar}\int d\tau\Psi_{j}^{\prime\ast
}(t)[H,V(t)]\Psi_{j}^{\prime}(t)=0. \label{efn}%
\end{equation}
The time-averaged power of the external force is zero. Thus the change in
kinetic energy is due to the internal force:%

\begin{equation}
\frac{1}{T}\int_{0}^{T}dt\frac{d}{dt}K_{\Psi_{j}^{\prime}}(t)
=\frac{1}{T}\int_{0}^{T}dt\frac{1}{i\hbar}\int d\tau\psi_{j}^{\prime\ast
}(t)[K_{0},H]\psi_{j}^{\prime}(t)\label{inft}%
\end{equation}%
\[
=\frac{1}{i\hbar}\sum_{m(\neq j)}\{K_{jm}\sum_{k(\neq j)}[\frac{F_{mk}%
F_{jk}^{\ast}}{\hbar(\omega_{kj}+\omega)}+\frac{F_{km}^{\ast}F_{kj}}%
{\hbar(\omega_{kj}-\omega)}]
+\frac{(F_{mj}F_{jj}^{\ast}-F_{jm}^{\ast}F_{jj})K_{jm}}{\hbar\omega}%
\]%
\[
-K_{mj}\sum_{k(\neq j)}[\frac{F_{mk}^{\ast}F_{jk}}{\hbar(\omega_{kj}+\omega
)}+\frac{F_{km}F_{kj}^{\ast}}{\hbar(\omega_{kj}-\omega)}]
-\frac{(F_{mj}^{\ast}F_{jj}-F_{jm}F_{jj}^{\ast})K_{mj}}{\hbar\omega}\}
\]%
\[
+\frac{1}{i\hbar}\sum_{kl(k\neq l)}K_{lk}(\varepsilon_{k}-\varepsilon
_{l})\{\frac{F_{kj}F_{lj}^{\ast}}{\hbar(\omega_{kj}-\omega)\hbar(\omega
_{lj}-\omega)}
+\frac{F_{jk}^{\ast}F_{jl}}{\hbar(\omega_{kj}+\omega)\hbar(\omega_{lj}%
+\omega)}\}.
\]
For an AC voltage, because $[K_{0},H]=[K_{0},H_{2}]$, we may say that the
Joule heat comes from the power of the internal force. For $\omega=0$,
Eq.(\ref{inft}) reduces to Eq.(\ref{inw1}), the Joule heat for DC voltage.

\section{Electromagnetic field}

\label{em}

In Sec.\ref{si} and \ref{psi}, a special gauge is used: both DC voltage and
low frequency AC voltage are described by scalar potentials. The rates of
change kinetic energy are given in (\ref{kcs},\ref{en}). Now consider the
system interacting with a general electromagnetic field described by
($\mathbf{A},\phi$) which may or may not change with time. We will not
restrict ourselves to any special gauge. The kinetic energy operator of the system is
$\widehat{K}(t)$ rather than $K_{0}$. If
vector potential $\mathbf{A}$ changes with time, there is one more term
$\partial\widehat{K}(t)/\partial t$ in $dK_{\Psi_{j}^{\prime}}(t)/dt$ (\ref{rat}). $\partial\widehat{K}(t)/\partial t$
results from the time dependence of vector potential $\partial\mathbf{A}%
(\mathbf{r},t)/\partial t$.

To apply TDPT to compute $dK_{\Psi_{j}^{\prime}}(t)/dt$, we notice that
$H_{mf}=V_{A}+V_{\phi}$, where $V_{A}=V_{A1}+V_{A2}$ is the interaction
involving vector potential $\mathbf{A}$, $V_{A1}$ presents the terms which are
first order in $\mathbf{A}$, $V_{A2}$ presents the terms which are second
order in $\mathbf{A}$. $V_{A2}$ is only a function of coordinates and does not
include differential operators. Notice $\widehat{K}=K_{0}+V_{A}$ and
$H=K_{0}+H_{2}$, the commutator in the first term of (\ref{rat}) can be
transformed to%
\begin{equation}
\lbrack\widehat{K},H^{\prime}]=[\widehat{K},H]+[H,H_{mf}]+[V_{A1},V],
\label{nd}%
\end{equation}
where $V=V_{\phi}+H_{2}$ and
\begin{equation}
\frac{1}{i\hbar}[V_{A1},V]=\sum_{i}m^{-1}[e\mathbf{A}(\mathbf{r}_{i}%
,t)\cdot\nabla_{\mathbf{r}_{i}}V]
-\sum_{\alpha}M_{\alpha}^{-1}[Z_{\alpha}e\mathbf{A}(\mathbf{R}_{\alpha
},t)\cdot\nabla_{\mathbf{R}_{\alpha}}V].\label{dfn}%
\end{equation}%
Since $-\nabla_{\mathbf{r}_{i}}V=\mathbf{f}_{i}+[-e\nabla_{\mathbf{r}_{i}}%
\phi(\mathbf{r}_{i},t)]$, $-e\mathbf{A}(\mathbf{r}_{i},t)/m$ is the part of
velocity due to field of the $i^{th}$ electron, the first term in (\ref{dfn})
is the power of the field momentum due to scalar potential $\phi$ and the
internal force. The 2$^{nd}$ term is the power of the field momentum of nuclei.
Substituting (\ref{nd}) into (\ref{rat}), one has%
\begin{equation}
\frac{d}{dt}K_{\Psi_{j}^{\prime}}(t)=\frac{1}{i\hbar}\int d\tau\Psi
_{j}^{\prime\ast}(t)([\widehat{K},H]+[H,H_{mf}(t)])\Psi_{j}^{\prime}(t)
\label{rat1}%
\end{equation}%
\[
+\int d\tau\Psi_{j}^{\prime\ast}(t)(\frac{\partial\widehat{K}(t)}{\partial
t}+\frac{1}{i\hbar}[V_{A1},V])\Psi_{j}^{\prime}(t).
\]
For an electromagnetic field with several frequencies $\omega_{n}$, the matrix
element of the interaction has form:
\begin{equation}
\lbrack H_{mf}(t)]_{lj}=\sum_{n}[F_{lj}^{n}e^{-i\omega_{n}t}+F_{jl}^{n\ast
}e^{i\omega_{n}t}]. \label{gem}%
\end{equation}
With the same method for an AC voltage described by a scalar potential, one
can show to 2$^{nd}$ order in $H_{mf}(t)$, $(i\hbar)^{-1}\int d\tau
\Psi^{\prime\ast}(t)[H,H_{mf}]\Psi^{\prime}(t)=0$. The first term in
(\ref{rat1}) can be similarly obtained as (\ref{inft}) for AC voltage:%
\begin{equation}
\frac{1}{i\hbar}\int d\tau\Psi_{j}^{\prime\ast}(t)[\widehat{K},H]\Psi
_{j}^{\prime}(t)=
\frac{1}{i\hbar}\sum_{p=1,2}\{\sum_{l(\neq j)}(\varepsilon_{j}-\varepsilon
_{l})[a_{lj}^{(p)\ast}(t)e^{it(\varepsilon_{l}-\varepsilon_{j})/\hbar}K_{lj}\label{win}
\end{equation}%
\[
-a_{lj}^{(p)}(t)e^{it(\varepsilon_{j}-\varepsilon_{l})/\hbar}K_{jl}]\}
+\frac{1}{i\hbar}\sum_{kl(k\neq l)}a_{lj}^{(1)\ast}(t)a_{kj}^{(1)}%
(t)e^{it(\varepsilon_{l}-\varepsilon_{k})/\hbar}K_{lk}(\varepsilon
_{k}-\varepsilon_{l}),
\]
where the matrix elements of the kinetic energy are calculated with
$\widehat{K}(t)$ rather than $K_{0}$, the transition probability amplitudes
are computed for $H_{mf}(t)$. After average over one period of external
field, the order V term in the curly bracket is zero. We see from (\ref{pkpt}) that
$\partial\widehat{K}(t)/\partial t$ is first order in vector potential. To
second order in $H_{mf}(t)$,
\begin{equation}
\int d\tau\Psi_{j}^{\prime\ast}(t)\frac{\partial\widehat{K}(t)}{\partial
t}\Psi_{j}^{\prime}(t)=\int d\tau\Psi_{j}^{\ast}\frac{\partial\widehat{K}%
(t)}{\partial t}\Psi_{j} \label{kta}%
\end{equation}%
\[
+\sum_{l}[a_{lj}^{(1)\ast}(t)e^{it(\varepsilon_{l}-\varepsilon_{j})/\hbar}\int
d\tau^{\prime}\Psi_{l}^{\ast}\frac{\partial\widehat{K}(t)}{\partial t}\Psi_{j}
+a_{lj}^{(1)}(t)e^{-it(\varepsilon_{l}-\varepsilon_{j})/\hbar}\int d\tau
\Psi_{j}^{\ast}\frac{\partial\widehat{K}(t)}{\partial t}\Psi_{l}].
\]
For many choices of gauge, $\partial\widehat{K}(t)/\partial t$ is Hermitian,
the second term in the square bracket is the complex conjugate of the first term.
Combining (\ref{dfn},\ref{win},\ref{kta}), the rate $dK_{\Psi_{j}^{\prime}%
}(t)/dt$ of change in kinetic energy in (\ref{rat1}) is determined.

We analyze the conditions which lead to KGF for this general case. If (i) the gradient of the
carrier density is small; and (ii) the wavelength of vector potential is
larger than the characteristic length of the considered sample, the first term
in (\ref{kta}) is zero. One can see this from (\ref{pkpt}): under conditions
(i) and (ii), the second term in (\ref{pkpt}) is ignored, and the first term
in (\ref{kta}) becomes $\left[  -\partial\mathbf{A}(\mathbf{r};t)/\partial
t\right]  \sum_{i}\int d\tau\Psi_{j}^{\ast}\mathbf{v}_{i}\Psi_{j}$. But the
average velocity $\int d\tau\Psi_{j}^{\ast}\mathbf{v}_{i}\Psi_{j}$ in a
stationary state $\Psi_{j}$ of $H$ is zero. Now only the square bracket term is
left in (\ref{kta}). Each term represents the power due to part of electric
field described by vector potential:%
\begin{equation}
-\partial\mathbf{A}(\mathbf{r};t)/\partial t=-\partial\mathbf{A}_{\perp
}(\mathbf{r};t)/\partial t-\partial\mathbf{A}_{\parallel}(\mathbf{r}%
;t)/\partial t, \label{vef}%
\end{equation}
the longitudinal and transverse parts satisfy $\nabla\times\mathbf{A}%
_{\parallel}=0$ and $\nabla\cdot\mathbf{A}_{\perp}=0$ respectively. The
contribution from $-\nabla\phi$ is absent from (\ref{kta}).

Greenwood used a special gauge: $\mathbf{A}(t)=-t\mathbf{E}$ and $\phi=0$ to
describe a DC voltage\cite{Gre}. Then $V_{A1}=t\mathbf{E}\cdot\mathbf{v}$,
according to TDPT, $\langle l|e\mathbf{E}t\cdot\mathbf{v}|j\rangle
\thicksim(\varepsilon_{l}-\varepsilon_{j})a_{lj}^{(1)}/t$. If the interaction
time is long enough that transition probability per unit time is well defined,
then a typical term in the square bracket of (\ref{kta}) becomes $\sum_{l}(\varepsilon
_{l}-\varepsilon_{j})|a_{lj}^{(1)}|^{2}/t$. Averaging (\ref{kta}) over the
occupation probability of the initial state $\Psi_{j}$, one obtains
(\ref{dcc}). From (\ref{rk1},\ref{trc1}) and (\ref{rat1}), (\ref{dcc}) missed
the contributions from (\ref{dfn}) and (\ref{win}). Although the momentum due
to field i.e. (\ref{dfn}) is negligible, the contribution (\ref{win}) is same
order as (\ref{kta}): $|(V_{A1})_{lj}|^{2}/\hbar$.

If we consider only the radiation field in (\ref{vef}), $\partial
\mathbf{A}_{\perp}(\mathbf{r};t)/\partial t\thicksim\omega A_{\perp}$. Then
$a_{lj}^{(1)}(t)=-\langle l|e\mathbf{A}_{\perp}\cdot\mathbf{v}|j\rangle
e^{i(\omega_{lj}-\omega)t+\lambda t}[\hbar(\omega_{lj}-\omega-i\lambda)]^{-1}%
$. If the interaction time is long enough, the square bracket in (\ref{kta})
becomes $\sum_{l}\hbar\omega|a_{lj}^{(1)}(t)|^{2}/t$~ i.e. (\ref{abs}). For a
zero frequency radiation field, the absorbed energy from field is zero. Thus we
understand that although the limit procedure from (\ref{acc}) to (\ref{dcc})
is not legitimate, (\ref{dcc}) can be obtained in Greenwood gauge (a
longitudinal field), and contains part of the conductivity. The
power (\ref{win}) induced by the internal force on system always exists, no
matter longitudinal field or transverse field. The applicable lower limit
frequency of (\ref{acc}) is at least $\omega>>\sigma/\epsilon_{0}$. For
intrinsic Si\cite{ash}, $\sigma=1.2\times10^{-5}$Scm$^{-1}$, the simplified
version\cite{motda,mos} of Kubo-Greenwood formula wok only when the frequency
of external field is higher than $\sigma/\epsilon_{0}\thicksim1.4\times10^{8}$Hz.

Numerical caculations\cite{gal,fau,dha,bin,hub} have shown 
$\sigma(\omega)$ approaches zero with $\omega$
below a resonance absorption
frequency $\omega_{r}$ which is roughly the splitting from the maximum of
density of states in the valence band to the edge of conduction band. This agrees with our
analysis. Sometimes researchers extropolate\cite{gal} $\sigma_{dc}$ from
$\sigma(\omega_{r})$. Because of the existence of (\ref{dcc}), if the velocity
matrix elements are not very sensitive to the energies of states, extrapolation
will deliver reasonable results.

\section{Conclusion}

\label{cosn}

In summary, from the rate of change in kinetic energy, we obtained a rigorous
expression for the power done by an arbitrary electromagnetic field in any
gauge. It leads to a proper current density which has been proved by the
continuity equation\cite{short}, polarization density\cite{4c} and current
operator\cite{sg}. We show the simplified derivation of KGF by Mott-Davis and
Moseley-Lukes suffers the same approximations used by Greenwood. The work done
by the internal force is missed in (\ref{acc},\ref{dcc}), they are same order
as the terms in KGF. Using (\ref{dcc}) or extrapolation from (\ref{acc}) can obtain a significant part of the DC conductivity, but a stricter calculation based on the rigorous current density would deliver more accurate, possibly even qualitatively new results.
\ack We thank the Army Research Office for support under MURI
W91NF-06-2-0026, and the National Science Foundation for support under
grant DMR 09-03225.


\begin{thebibliography}{99}                                                                                               %

\bibitem {motda}N. F. Mott and E. A. Davis, \textit{Electronic Processes in
Non-crystalline Materials}, Clarendon Press, Oxford, (1971).

\bibitem {gal}G.Gali, R. M. Martin, R. Car and M. Parrinello, Phys. Rev. Lett.
\textbf{63}, 988 (1989); Phys. Rev. B \textbf{42}, 7470 (1990).

\bibitem {Gre}D. A. Greenwood, Proc. Phys. Soc. (London) \textbf{71}, 585 (1958).

\bibitem {sg}M.-L. Zhang and D. A. Drabold, arXiv: 1011.1527, submitted to
Phys. Rev. E.

\bibitem {mos}L. L. Moseley and T. Lukes, Am. J. Phys. \textbf{46}, 676 (1978).

\bibitem {abt}T. A. Abtew, M.-L. Zhang and D. A. Drabold, Phys. Rev. B
\textbf{76}, 045212 (2007).

\bibitem {short}M.-L. Zhang and D. A. Drabold, Phys. Rev. Lett. \textbf{105},
186602 (2010).

\bibitem {4c}M.-L. Zhang and D. A. Drabold, arXiv: 1008.1067, submitted to
Phys. Rev. B.

\bibitem {lv3}L. D. Landau and E. M. Lifshitz, Quantum Mechanics, 3rd edition,
Pergamon Press, Oxford (1977).

\bibitem {lv8}L. D. Landau, E. M. Lifshitz and L. P. Pitaevski\u{\i},
Eletrodynamics of Continuous Media, 2nd edition, Butterworth Heinemann Ltd,
Oxford (1984).

\bibitem {ash}N. W. Ashcroft and N. D. Mermin, Solid State Physics, Saunders
College Publishing,\ Fort Worth (1976).



\bibitem {fau}G. Faussuriera, C. Blancarda, P. Renaudina and P.L.
Silvestrellib, Journal of Quantitative Spectroscopy and Radiative Transfer
\textbf{99}, 153 (2006).

\bibitem {dha}M. W. C. Dharma-wardana, Phys. Rev. E \textbf{73}, 036401 (2006).

\bibitem {bin}B. Cai and D. A. Drabold, \textquotedblleft AC conductivity in
Ge$_{2}$Te$_{2}$Sb$_{5}$\textquotedblright, in preparation.

\bibitem {hub}A. H\"{u}bsch, R. G. Endres, D. L. Cox, and R. R. P. Singh,
Phys. Rev. Lett. \textbf{94}, 178102 (2005).
\end{thebibliography}
\end{document}